\documentstyle{article}
\begin{document}
\title{Covariant Perturbation  Theory of\\  Non-Abelian Kinetic Theory
}
\author{{Zheng Xiaoping$^{1 \hskip 2mm 2}$\thanks{Email address:
  zhengxp@iopp.ccnu.edu.cn}\ \  Li Jiarong$^{1}$}\\
{\small 1 The Institute of Particle Physics, Huazhong Normal
University, Wuhan, P.R.China}\\
{\small 2 Department of Physics, Zhongshan University, Guangzhou, P.R.China}
}
\maketitle

\begin{abstract}
\vskip0.5cm
\begin{minipage}{120mm}
\hskip0.8cm A new 'double perturbation' theory is presented in the framework 
of the kinetic theory of quark-gluon plasma. A solvable set of equations from 
the 'double perturbation' is derived and are shown to be gauge-invariant. 
As an example, the Landau damping rate for the plasmon at zero momentum 
is calculated and discussed.

\end{minipage}
\end{abstract}
\vskip 0.9cm

In the last decade, much interest was focused on the study of non-Abelian
plasma from transport theory~\cite{r1}. It is generally believed that  
kinetic theory can describe correctly the quark-gluon plasma (QGP) physics 
just as what can be done by the temperature field theory~\cite{r3}, 
and is ready to be extended to out-of-equilibrium situations~\cite{r4}.  
It has been demonstrated that the hard thermal loop (HTL) in temperature 
field theory can be obtained from the QGP kinetic theory~\cite{r2,r5}, 
but up to now there does not exist a valid scheme for solving the kinetic 
equations and at the same time keeping the non-Abelian gauge symmetry.

A standard perturbative approach to the  kinetic equations of plasma,
which can keep the guage symmetry, is to expand 
the equations in the ascending power of guage coupling constant~\cite{r2}. 
Another popular method in traditional study of plasma is to expand the 
equations in powers of weak field strength~\cite{r4,r6,r7}. Since
these two methods are in full agreement with each other in the study of
electromagnetic plasma~\cite{r8,r9} due to the Abelian nature of the
dynamics (linear dynamics), they have been widely used 
in the literatures for QGP. However, they both suffer from powerlessly
overcome
shortcomings: The former, as is well known,
gives only a linear leading order, i.e. an Abelian-like contribution, 
instead
of truly  non-Abelian ones, while the latter breaks the non-Abelian gauge 
covariance badly as was pointed out in Ref.~\cite{r2,r7}. 

If we want to solve the above-mentioned problem in the
framework of kinetic theory, we must pay 
special attention to the following two aspects: the preserve of gauge
invariance in the perturbation process
and the treatment of the nonlinearity in the
equations because of the nonlinear (non-Abelian) nature of QGP.   

In the present letter, we will propose a new scheme both satisfying the SU(3) 
gauge symmetry and having solvability for  non-Abelian  kinetic theory. 
The basic idea of the scheme is:
After  expanding  the kinetic equations in guage coupling constant $g$, 
do the iterative calculation in powers of field strength for the purely 
non-Abelian counterpart. In the following we will call this scheme
as 'double perturbation'. 

Our aim is to provide a well-defined prescription for treating the 
color dielectric physics.  In this letter, we will first present  the 
framework of the new perturbation theory and then give the main
results of our analysis. 
This includes the derivation of a general set of perturbative
kinetic equations  for non-Abelian plasma and the comments of the 
gauge invariance and solvability of these equations.  An application 
of the new formalism in the explicit calculation of Landau damping rate in 
close-to-equilibrium QGP, which has been computed from HTL in field 
theory~\cite{r10,r11}, will be given as an example. 
Finally, we will show some evidence that our results are associated with or 
even go beyond the HTL approximation.
A more extensive discussion and further details on computations will be 
given in a longer article~\cite{r12}.

Let us begin with the derivation of a set of equations describing the
dynamics associated with non-Abelian fluctuations. It is sufficient to
adopt the semiclassical kinetic theory of QGP for studying the thermal
effects. The theory  consists of kinetic equations and field equation.
The kinetic equations  read as~\cite{r13}
\begin{equation}
p^\mu D_\mu Q_{\pm}({\bf p},x)\pm {g\over
2}p^\mu\partial^\nu_p\{F_{\mu\nu}(x),Q_\pm ({\bf p},x)\}=0,
\end{equation}
\begin{equation}
p^\mu \tilde D_\mu G({\bf p},x)+{g\over 2}p^\mu\partial^\nu_p\{{\tilde
F}_{\mu\nu}(x),G({\bf
p},x)\}=0,
\end{equation}
where the letters with \~{} represent the corresponding operators
in adjoint representation of SU(N), in which the generator is $T_a$.

The background field equation is written as
\begin{equation}
D^\mu F^a_{\mu\nu}=j^a_\nu(x),
\end{equation}
where covariant derivative $D^\mu=\partial^\mu+igA^\mu(x)$, 
field potential $A^\mu\equiv A^\mu_a\tau^a$, and field tensor
$F^{\mu\nu}=\partial^\mu A^\nu-\partial^\nu A^\mu-igf_{abc}\tau^c
A^\mu_aA^\nu_b$. $\tau_a$ is the generator of SU(N). The color currents
including external and induced ones are $j^a_\nu=j^{{\rm ext}\ a}_\nu+
j^{{\rm ind}\ a}_\nu$.
 
In principle, a perturbative method is often necessary to solve nonlinear
equations. Here, for simplicity, we take the close-to-equilibrium situation 
of a QGP as an example to discuss this method. 
We will show later that our method is easily extended to
out-of-equilibrium QGP. 

\def\la{\langle} \def\ra{\rangle}
Following the same philosophy in traditional kinetic theory we write the
distribution functions $Q_{\pm}(x,p), G(x,p)$ and fields as~\cite{r2,r13},
\begin{equation}
A^a_\mu=a^a_\mu,\ \ \ Q_{\pm}=Q^{\rm eq}_\pm +\delta Q_\pm ,\ \
G=G^{\rm eq}+\delta G.
\end{equation}
The associated density fluctuations and induced fields are random
quantities, satisfying  $\la \delta Q\ra =0,\la \delta G\ra=0$ 
and $\la a\ra =0$, where
symbol $\la \ \ra $ denotes the random-phase average. We
define the field tensor corresponding to $a_\mu$ as $f_{\mu\nu}$.

In this way we
obtain a set of kinetic equations for fluctuations from equations (1) and (2)
\begin{equation}
p^\mu D_\mu\delta Q_{\pm}({\bf p},x)\pm {g\over
2}p^\mu\partial^\nu_p\{f_{\mu\nu}(x),Q_\pm ({\bf p},x)\}=0,
\end{equation}
\begin{equation}
p^\mu {\tilde D}_\mu\delta G({\bf p},x)+{g\over 2}p^\mu\partial^\nu_p\{\tilde
{f}_{\mu\nu}(x),G({\bf
p},x)\}=0.
\end{equation}
 
In order to ensure the consistency of soft covariant derivative,
in the following we will 
impose a limitation on the amplitude of the fields: if  
 $a_\mu\sim T$, then $ga_\mu\sim gT$ is of the same order as the derivative
of a slowly varying quantity, $i\partial_\mu\sim ga_\mu$~\cite{r14}.  

We now employ our 'double perturbation' scheme to 
expand the distribution functions in powers of $g$ in the first step, and
then iterate repeatedly the nonlinear parts in the expanded equations in field
quantity. These can be expressed as
\begin{equation}
\delta Q=gQ^{(1)}+g^2Q^{(2)}+\cdots,\ \
\delta G=gG^{(1)}+g^2G^{(2)}+\cdots,
\end{equation}
\begin{equation}
Q^{(n)}=\sum\limits_{\lambda=1} Q^{(n,\lambda)}, \ \
G^{(n)}=\sum\limits_{\lambda=1}
G^{(n,\lambda)}.
\end{equation}
Thus we obtain a series of equations
\begin{eqnarray}
p^\mu\partial_\mu Q^{(n)}_\pm\pm ig\sum\limits_\lambda p^\mu[a_\mu,
Q^{(n,\lambda)}_\pm]+{g\over 2} p^\mu\{f_{\mu\nu},
\partial^\nu_p Q^{(n-1)}_\pm\}
=0,
\end{eqnarray}
\begin{eqnarray}
p^\mu\partial_\mu G^{(n)}+ ig\sum\limits_\lambda p^\mu[\tilde{ a}_\mu,
G^{(n,\lambda)}]+{g\over 2} p^\mu\{\tilde{ f}_{\mu\nu},
\partial^\nu_p G^{(n-1)}\}
=0,
\end{eqnarray}
\begin{equation}
D^\mu(a) f^{(n)a}_{\mu\nu}=j^{(n){\rm ind}\ a}_\nu(x),
\end{equation}
where $n$ and $\lambda$ represent the powers of coupling constant and 
field quantity, respectively. The induced currents associated with each 
order of density fluctuations are determined by 
\begin{equation}
j^{(n){\rm ind}\ a}_\nu=g\int{d^3p\over (2\pi)^3}{p_\nu\over p_0}
{\rm Tr}\left (2N_f\tau^a[Q_+^{(n)}-Q_-^{(n)}]+2T^aG^{(n)}\right )
\end{equation}
where $N_f$ is the number of quark flavour. The factor 2 
accounts for the spin degrees of freedom. 

Equations (9), (10), (11) and (12) form a basis of  perturbation theory of
non-Abelian kinetic theory. 

Some short comments follow.

1. The non-Abelian gauge symmetry is exactly preserved in the
perturbation equations (9) and (10) for each order of $g$. 
As a consequence $Q^{(n)}_\pm$ and $G^{(n)}$, like $Q_\pm$ and $G$, 
transform separately as SU(3) gauge-invariant scalars. 
Because the summation of non-Abelian terms in (9)
and (10) can be done infinitely, the 'double perturbation' method we use
guarantees the gauge-invariance of the result of a physical quantity 
within any high precision ($\lambda$ being an arbitrarily large number). 

2. In general, it is a difficult task to straightforwardly solve the equations
(5) and (6) due to the the nonlinearity of non-Abelian counterparts~\cite{r15}. 
However, the summation over $\lambda$ in equations (9) and
(10) imply that the iterative procedure in the powers of field quantity
ensures the solvability of the non-Abelian counterparts. 

3. As was pointed out in Ref.~\cite{r16},
the perturbative expansion in field quantity only, 
which had been carried out by some authors~\cite{r4,r6,r7}, have to suffer 
from the disadvantage of breaking non-Abelian gauge symmetry at 
each step of approximation calculation. This is because the results of
all higher orders in field quantity contain the contributions from the
relevant lower orders in coupling constant $g$ for the density fluctuations.
So, in order to get a gauge-invariant physical result,
the problem of resummation of all the contributions of the same order 
in $g$ has to be taken into account in the past works~\cite{r6,r7,r16}. 
While our approach 
automatically singles out all the contributions of the same order in $g$
and collects them together. In this sense, the
treatment of non-Abelian contributions in our approach gives a 
theoretical basis for the resummation technique.
   
4. It can be easily verified from Eq.(12) that not only the total color 
current but also each order current obey the covariant conservation law
\begin{equation}
D^\mu j^{(n){\rm ind}\ a}_\mu =0.
\end{equation}
This is automatically consistent with field equation (11).

5. Our theory can be easily generalized to the out-of-equilibrium situations 
through the decomposion of the distribution functions $Q_\pm (x,p), G(x,p)$ 
and fields $A_\mu$ into regular terms and density fluctuations. The
expression (4) is then replaced by
\begin{equation}
A^a_\mu=\la A_\mu^a\ra + a^a_\mu,\ \ \ Q_{\pm}=\la Q_\pm\ra  +\delta Q_\pm ,\ \
G=\la G\ra+\delta G.
\end{equation}
Such kind of division has been used in the study of QGP~\cite{r4} as well as 
in other works~\cite{r7,r15,r16}. In particular,
the gauge consistency of the decomposition has been proved in 
Ref.~\cite{r15}. Our stress here is to put forward a gauge-consistent
fluctuation dynamics($a_\mu\sim T$), while Litim et al think that the regular parts
describe mean field dynamics~\cite{r15} in full
accordance with the effective soft field dynamics($\la A_\mu\ra\sim gT$)~\cite{r17}.  

6. The 'double perturbation' approach gives an insight into  the plasma with 
color freedom of degrees. 

We will get a more penetrating understanding of these remarks from the
calculations below.

As an example, we now apply the above formalism to calculate the gloun
damping rate for a purely gluonic gas in the close-to-equilibrium situation. 
The first order equation in coupling constant is 
\begin{eqnarray}
p^\mu\partial_\mu G^{(1)} + ig\sum\limits_\lambda p^\mu[\tilde{ a}_\mu,
G^{(1,\lambda)}]+gp^\mu\tilde{f}^{(1)}_{\mu\nu}\partial^\nu_p G^{(0)}
=0
\end{eqnarray}
\begin{equation}
D^\mu(a) f^{(1)a}_{\mu\nu}=j^{(1){\rm ind}a}_\nu(x),
\end{equation}
Denoting the summation term by $\tilde {S}$, we have
\begin{eqnarray}
\tilde {S}=&-&\int{d^4k\over (2\pi)^4}\partial_p^\nu G^{(0)}
\left ( ig\int {d^4k_1\over (2\pi)^4}{d^4k_2\over (2\pi)^4}\delta (k-k_1-k_2)
{1\over p\cdot k_2} 
\right.\nonumber
[p\cdot \tilde{a}(k_1), p^\mu \tilde{f}^{(1)}_{\mu\nu}(k_2)] \nonumber\\
&+&ig^2\int {d^4k_1\over (2\pi)^4}{d^4k_2\over (2\pi)^4}{d^4k_3\over (2\pi)^4}
\delta (k-k_1-k_2-k_3)
{1\over p\cdot (k_2+k_3)p\cdot k_3}\nonumber\\
&\ &\hskip2cm \times 
[p\cdot \tilde{a}(k_1), 
[p\cdot \tilde{a}(k_2), p^\mu \tilde{f}^{(1)}_{\mu\nu}(k_3)]]\nonumber\\
&+&\cdots\cdots \biggr )
\end{eqnarray}

It is worth noticing that all the terms in $\tilde {S}$ are of same order
in $g$. This is because the factor $g$ in the numerator and the 4-dimensional
wave vector $k_\mu$ $(\sim \partial_\mu \sim g)$ in the denominator appear
in pairs  and cancell the factor $g$ each other. The remaining dependency on $g$,
coming from the field strength $\tilde{f}^{(1)}_{\mu\nu}$, is of  first order.

A general formula of $G^{(1)}(\omega,{\bf k})$ in the momentum space can be
written down from Eq.(15).
We know that the current induced by the fluctuations can be expressed as
\begin{eqnarray}
j^{(1){\rm ind}\ a}_\nu=g\int {{\rm d^3p}\over (2\pi)^3}{p_\nu\over p_0}{\rm
Tr} [2T_aG^{(1)}({\bf p},x)]\},
\end{eqnarray}

Following the same method in~\cite{r4}, we can derive the response equation 
of medium from kinetic and field equations. As an example, we do this 
 approximately by taking  $\lambda $ up to 2 and consider the case of zero momentum.
We get from field equation (16) together with equations (15) and (18),  
{\small
\begin{eqnarray}
&\null{}&-\omega^2\varepsilon^{(\sigma)}(\omega,0)\la a^2(\omega,0)\ra \nonumber\\
&\null{}&\hskip0.8cm =Ng^2\int {d\omega_1d^3k_1\over (2\pi)^4}
d\omega_2\delta(\omega-\omega_1-\omega_2){1\over\omega^2\varepsilon^{(\sigma)}}\nonumber\\
&\null{}&\hskip2cm \times (\kappa_{ll}\la a^2_l(\omega_1,{\bf
k_1})\ra \la a^2_l(\omega_2,{\bf
k_1)}\ra +\kappa_{tl}\la a^2_t(\omega_1,{\bf
k_1)}\ra \la a^2_l(\omega_2,{\bf
k_1})\ra\nonumber\\
&\null{}&\hskip2.8cm+\kappa_{tt}\la a^2_t(\omega_1,{\bf
k_1})\ra \la a^2_t(\omega_2,{\bf
k_1})\ra )\nonumber\\
&\null{}&\hskip0.8cm+Ng^2\int {d^4k_1\over(2\pi)^4}(\la {\bf a}^2(k_1)\ra -
\la a ^2(k_1)\ra )\la a^2(\omega, 0)\ra \nonumber\\
&\null{}&\hskip0.8cm+Ng^2m_g^2\int {d^4k_1\over(2\pi)^4}d{\bf v}{1\over\omega}{1\over
v\cdot(k_1-k)}\left({\omega_1\over v\cdot k_1}-1\right)
\la ({\bf v\cdot a})^2(k_1)\ra \la({\bf
v\cdot a})^2(\omega,0)\ra = 0,
\end{eqnarray}
}
where $\varepsilon^{(\sigma)}$ denotes the dielectric function for $\sigma$
mode, $\sigma=l$ or $t$ represent the longitudinal or transverse wave, 
respectively.
$\kappa_{ll},\kappa_{tl},\kappa_{tt}$ are coefficient functions of
$\omega_1,\omega_2, {\bf k}_1$, associated with the definite interactions of
plasmons.

We know that damping is connected with the imaginary parts in Eq.(19).
Obviously, the second term of the right-hand side has no imaginary part.
The damping will originate from the physical processes described by the 
first and third terms. Therefore, the Landau damping rate can be obtained
easily:
\begin{equation}
\gamma^{(\sigma)}(0)=(a^s+a^c){Ng^2T\over 24\pi},
\end{equation}
with 
$$ a^s=6\int{\bf
k}^2_1dk_1d\omega_1d\omega_2\delta(m_g-\omega_1-\omega_2)(K_{ll}\rho_l\rho_l+K_{tl}\rho_t\rho_l+K_{tt}\rho_t\rho_t)$$
$$ a^c=24\pi\int k^2_1dk_1d\omega_1({(\omega_1-m_g)^3\over
m_gk_1\omega_1^4}\rho_l(k_1)-{\omega_1-m_g\over
m_gk_1\omega_1}(1-{(\omega_1-m_g)^2\over k_1^2})\rho_t(k_1)).$$
where  $K_{ll}={{\bf k_1}^4\over
\omega_1^3\omega_2^3}\kappa_{ll}, K_{tl}=-{{\bf k_1}^2\over
\omega_1\omega_2^3}\kappa_{tl}, K_{tt}={1\over\omega_1\omega_2}\kappa_{tt}$, 
$\rho_l$ and $\rho_t $ represent the longitudinal and transverse 
spectral densities, respectively. 

The forms of $a^s$ and $a^c$ show that two typical physical processes 
contribute to the Landau damping.  With the constraint 
of the $\delta$ function, The first one describes the self-coupling interaction of a mode,
which has a good formal correspondence  to the result of HTL in field 
theory~\cite{r11}. The other one represents the collisional interaction of
two plasmons or quasiparticles, which describes the long-range interaction
in QGP. The numerical result of $a^c$ has been obtained to be 5.973, while the
magnitude of $a^s$ is estimated to be of well-matched order with $a^c$.

In conclusion, we have proposed a 'double perturbation' approach and derived 
a series of perturbation equations for the non-Abelian kinetic theory. 
Our approach gives a new and deeper perspective to the perturbation theory 
and thereby provides a progress in the methodological problem in the study 
of non-Abelian kinetic theory.

The study of the dielectric physics in QGP will be advanced by applying this
new theory. This includes to get a non-Abelian gauge-consistent dielectric
tensor and discuss the response of a QGP to external sources~\cite{r4,r19},
and so on. We would emphatically point out that since our approach has led
$\lambda$-point functions (or correlators) into first-order fluctuations 
in coupling constant $g$, the physics beyond an equilibrium state will play 
more vital role in a QGP than in an electromagnetic plasma~\cite{r4,r7}. 

As an example, we have studied in the last part of the present letter 
the Landau damping in the zero momentum case using this approach.
The result shows that both the self-coupling interaction of a mode 
and the interaction of two plasmons contribute to the damping rate. The
physical mechanism for two-plasmon interaction can clearly be expressed as a
two-body 'collision' from long-range interaction of two quasiparticles. We
need to make it clear further as for the self-coupling interaction, which
has showed, in form, the similar 2-2 scattering and 2-3 scattering processes
contained in the self-energy of HTL~\cite{r11}.

We also believe that this approach will provide more and better results to
uncover the deeper link between kinetic theory and field theory 
searched  by many investigators~\cite{r2,r20}.

We thank Prof. Liu Lianshou for offering constructive comments and the help
in English. We also thank Chen Jisheng and Hou Defu for the useful
discussion. This work is supported by the National Nature Science Fund of China.

\end{document}